\documentclass[preprint,11pt]{aastex}
\usepackage{epsfig}

\newcommand{\km}{${\rm km\,s}^{-1}$}
\newcommand{\fuse}{{\em FUSE}}
\newcommand{\ovi}{O$\;${\small\rm VI}\relax}

\newcommand{\kms}{km~s$^{-1}$\relax}
\newcommand{\column}{cm$^{-2}$}

\let\la=\lesssim            
\let\ga=\gtrsim
\newcommand{\novi}{\ensuremath{N(\mbox{O$\,${\small VI}})}}
\newcommand{\N}[1]{\ensuremath{N(\mbox{#1})}}

\slugcomment{Draft of \today}
\shortauthors{Lehner \& Howk}
\shorttitle{Small-Scale structure of \ion{O}{6} toward NGC\,6752}
\begin{document}

\title{Small-Scale Structure of \ovi\ Interstellar Gas in the
  Direction of the Globular Cluster NGC 6752\altaffilmark{1}}

\altaffiltext{1}{Based on observations made with the NASA-CNES-CSA Far
Ultraviolet Spectroscopic Explorer. FUSE is operated for NASA by the
Johns Hopkins University under NASA contract NAS5-32985.}

\author{N.\ Lehner}
\affil{Department of Astronomy, University of Wisconsin, 
475 North Charter Street, Madison, WI 53706}
\email{nl@astro.wisc.edu}

\and

\author{J.\ C.\ Howk}
\affil{Center for Astrophysics and Space Science, 
University of California, San Diego, C-0424, La Jolla, CA 92093}
\email{howk@ucsd.edu}

\begin{abstract}
  In order to study the small-scale structure of the hot interstellar
  gas, we obtained {\em Far Ultraviolet Spectroscopic Explorer}\/
  interstellar \ovi\ interstellar absorption spectra of 4 four
  post-extreme horizontal branch stars in the globular cluster
  NGC\,6752 [$(l,b) = (336\fdg50,-25\fdg63)$, $d = 3.9$ kpc, $z= -1.7$
  kpc]. The good quality spectra of these stars allow us to measure
  both lines of the \ovi\ doublet at 1031.926 \AA\ and 1037.617 \AA.
  The close proximity of these stars permits us to probe the hot
  interstellar gas over angular scale of only $2\farcm2 - 8\farcm9$,
  corresponding to spatial scales $\la 2.5-10.1$ pc.  On these scales
  we detect no variations in the \ovi\ column density and velocity
  distribution. The average column density is $\log \langle N({\mbox
  \ovi})\rangle = 14.34 \pm 0.02$ ($\log \langle N_\perp ({\mbox
  \ovi})\rangle = 13.98$).  The measured velocity dispersions of the
  \ovi\ absorption are also indistinguishable.  Including the earlier
  results of Howk et al., this study suggests that interstellar \ovi\ 
  is smooth on scales $\Delta \theta \la 12\arcmin$, corresponding to
  a spatial scale of $\la 10$ pc, and quite patchy at larger scales.
  Although such small scales are only probed in a few directions, this
  suggests a characteristic size scale for the regions producing
  collisionally-ionized \ovi\ in the Galaxy.
\end{abstract}


\section{Introduction}

The phase structure of the interstellar medium (ISM) of galaxies is
ultimately determined by kinetic and radiative energy input from
stars.  The heating and ionization of the cold and warm phases of the
ISM are principally determined by the effects of stellar UV photons
both on atoms and dust grains and the interactions of cosmic rays with
the gas (see Wolfire et al. 1995).  The hotter phases of the ISM
ultimately owe their existence to the injection of energy into the ISM
by stellar winds and supernovae.  

Hot ($T\sim10^6$ K), X-ray emitting gas is produced in high-velocity
shocks, a process whose basic outline is understood.  The subsequent
evolution of hot material, including its cooling and interactions with
the cooler gas phases, is not well understood.  These processes can be
probed observationally using resonant absorption lines from
lithium-like ions \ion{C}{4}, \ion{N}{5}, and \ovi, which are produced
in collisionally-ionized gas at ``coronal temperatures'' ($T\sim10^5$
K; Sutherland \& Dopita 1993).  These ions have a range of ionization
energies, corresponding to a range of temperatures in equilibrium, and
all have resonance doublet transitions accessible to modern
ultraviolet observatories.  Of these, \ovi\ absorption is the most
important diagnostic to understanding gas in this temperature regime
since it cannot be produced via photoionization in the ISM of
galaxies, since the required pathlengths would be too large (see
Savage et al. 2003).

Measurements of \ovi\ absorption at high galactic latitudes have
accumulated in the recent years via observations with the {\em Far
Ultraviolet Spectroscopic Explorer}\/ ({\fuse}; see Wakker et al. 2003; 
Savage et al. 2003; Zsarg\`o et al. 2003, and references therein).  
Spitzer (1956) initially predicted a widespread distribution of hot gas in the
Galactic halo in order to confine neutral clouds seen at high
latitudes.  However, it is clear that the halo as seen in \ovi\ is not
the elegant, smooth halo in the idealized models of Spitzer.  Instead,
the distribution of \ovi, while grossly consistent with a thickened
layer of hot gas about the Milky Way's disk, is quite complicated
(Savage et al. 2003), with a wealth of structure on many scales.
Indeed, using \fuse\ observations of a series of closely-spaced
early-type stars in the Large and Small Magellanic Clouds (LMC and
SMC), \citet{howk02a} found large variations in \ovi\ column densities
over small angular scales, often finding $\ge50$\% variations in
$N(\mbox{\ovi})$ between pairs of stars separated by $ 0\fdg3 \la
\Delta \theta \la 5\fdg0 $.  This variation on small angular scale
extends to large angular scale \citep{savage00,savage03}.

However, the spatial/angular structure seen in \ovi\ column densities
is potentially providing us with a tool for understanding the physical
properties of the regions in which \ovi\ is produced.  In particular,
the scales of the structures seen in \ovi\ are likely indicating the
physical dimensions of typical \ovi\ regions, while an analysis of the
amount of change in $N(\mbox{\ovi})$ and the size scale of variations
could lead to information about the topology of the \ovi -bearing
structures.

The results of \citet{howk02a} have, to date, been the only thorough
analysis of the structure of \ovi\ on small scales.  They found column
density variations on even the smallest scales they were able to
probe, $\Delta \theta \sim 1\farcm8$, although their dataset included
only two pairs of sight lines with separations $<10\arcmin$.

\begin{figure}[tbp]
\epsscale{0.7}
\plotone{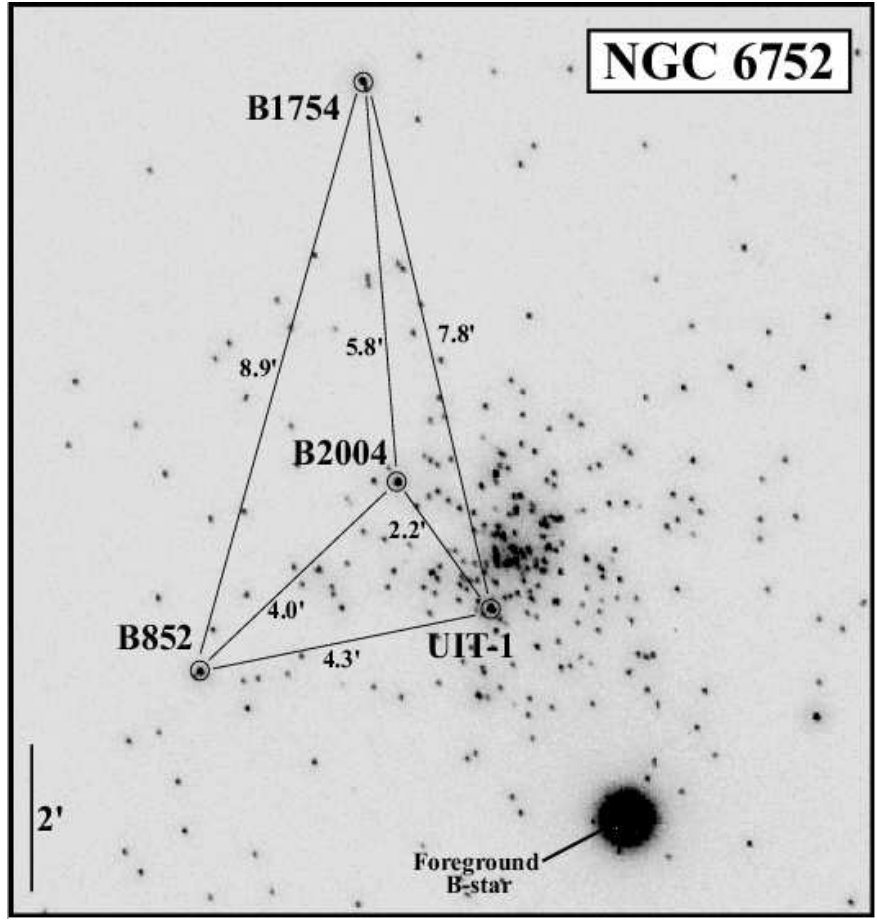}
\caption{The NUV ($\lambda_c = 1620$ \AA) UIT image of NGC\,6752 with 
  our background post-EHB stars marked.  The separation of these stars
  are marked, where $1\arcmin \simeq 1$ pc at the distance of 3.9 kpc
  of the globular cluster. All these stars were observed in the MRDS
  ($4\arcsec \times 20\arcsec$) aperture of {\em FUSE}\/ to exclude
  fainter stars that could contaminate our spectra.
\label{image}}
\end{figure}

To investigate the structure of \ovi\ on smaller angular scales, we
have obtained \fuse\ spectra of four post-extreme horizontal branch
(post-EHB) stars in the globular cluster NGC\,6752 [$(l,b) =
(336\fdg50,-25\fdg63)$, $d = 3.9$ kpc, $z= -1.7$ kpc, Harris 1996].
Figure~\ref{image} shows a NUV image ($\lambda_c = 1620$ \AA) of
NGC~6752 obtained with the {\em Ultraviolet Imaging Telescope} by
\citet{landsman96} during the Astro-2 mission of the Space Shuttle
{\em Endeavour} in 1995. We have marked our four post-EHB targets,
indicating the angular separations (in arcminutes) between each pair.
For reference $1\arcmin$ is about $1$ pc at the distance of NGC 6752.

The choice of stars in a single globular cluster allows us to probe
the degree of variation in the coronal temperature gas over small
angular ($2\farcm2 - 8\farcm9$) and spatial ($\la 2.5-10.1$ pc)
scales. Furthermore, the choice of hot, highly-evolved globular
cluster stars is important: such stars have proven to provide the
least complicated stellar continua in the regions near interstellar
\ovi\ \citep{howk03}, allowing very accurate column density
determinations. Indeed, population I OB-type stars, such as those
used as background sources in the LMC and SMC, can have quite complex
(and even temporally variable) continua in the regions near
interstellar \ovi\ due to their out-flowing stellar winds
\citep{lehner01,howk02b,hoopes02}.  The current study avoids this
source of uncertainty.

In the remainder of the paper, we will present the observations
(\S~\ref{obs}) and analysis (\S~\ref{ana}) of the \fuse\ \ovi\ data,
demonstrating that there is no detectable variability in \novi\ 
between our four sight lines.  In \S~\ref{disc}, we discuss the
implications of our measurements, and compare them with the previous
study of \citet{howk02a}. Lastly, we present concluding remarks in
\S~\ref{sum}.

\section{{\fuse}\/ Observations and Data Reduction}
\label{obs}

We obtained far-ultraviolet spectra of our four target post-EHB stars
using \fuse, which is described in detail by \citet{moos00,sahnow00},
in 2002 April and June as part of the GI program C076 (PI: Howk).  The
basic parameters of the observations are summarized in Table~\ref{t1},
including pertinent information about our target stars.  Each target
was observed using the MDRS ($4\arcsec \times 20\arcsec$) apertures
with the detectors in photon-address, or time-tag (TTAG), mode.  The
MDRS apertures were used with roll angle constraints to exclude other
stars in NGC 6752 that could potentially contaminate our spectra.  The
resolution of the data is $R\sim15,000$, or $\Delta v \sim 20$ \kms.

The {\fuse}\/ instrument has four channels: two optimized for the
short wavelengths (SiC\,1 and SiC\,2; 905--1100 \AA) and two optimized
for longer wavelengths (LiF\,1 and LiF\,2; 1000--1187 \AA). The
wavelength overlap of the various channels is important for
identifying potential fixed-pattern noise in the data. For three of
our datasets, only the LiF channels were aligned at the time of the
observations, providing us with coverage longward of 1000 \AA.  For
NGC\,6752-B852, the SiC channels were aligned as well, providing full
wavelength coverage.  The lack of full wavelength coverage for three
of our stars does not affect our scientific goals since the effective
areas of the LiF channels is much higher than the SiC channels at
\ovi\ (1031.926 and 1037.617 \AA).

\begin{figure}[tbp]
\epsscale{0.8}
\plotone{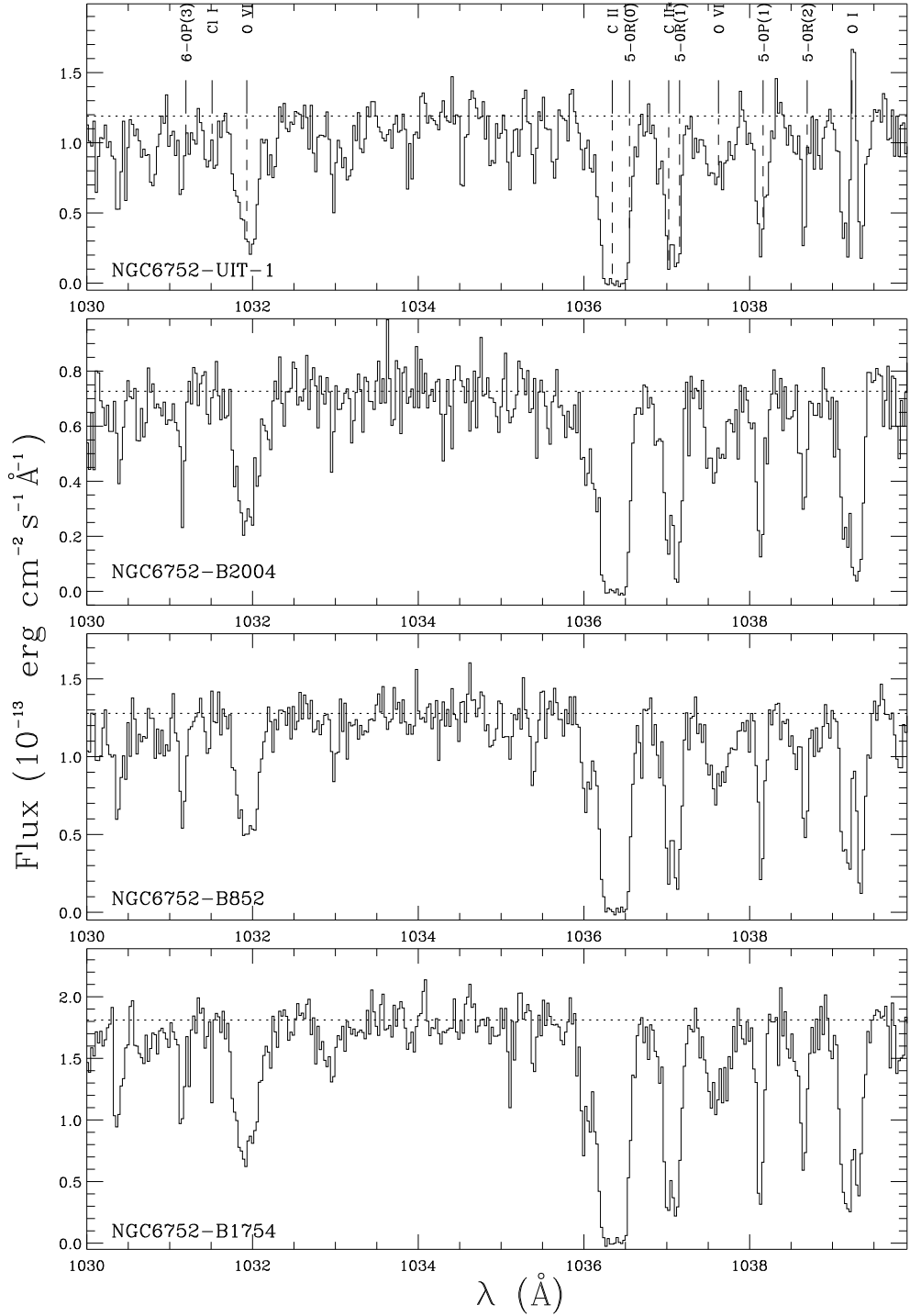}
\caption{Spectra of the different line of sights. The dotted line indicates the stellar continuum. 
  In the top diagram, we indicate the main interstellar (atomic,
  ionic, hydrogen molecular) absorption, and in particular the O\,{\sc
    vi} doublet.
\label{spec}}
\end{figure}

We used the standard calibration pipeline software (CalFUSE version
2.1.6) to extract and calibrate the spectra. The software screened the
data for valid photon events, removed burst events, corrected for
geometrical distortions, spectral motions, satellite orbital motions,
and detector background noise, and finally applied flux and wavelength
calibrations.  The extracted spectra associated with the separate
exposures were aligned by cross-correlating the positions of
interstellar absorption lines, and then co-added.  The co-added
spectra were finally rebinned by 4 pixels ($\sim 27$ m\AA) since the
extracted data are oversampled.  This provides approximately three
samples per 20 \km\ resolution element.  In Figure~\ref{spec} we show
the final spectra of our stars near the location of interstellar \ovi.

\begin{figure}[tbp]
\epsscale{1}
\plotone{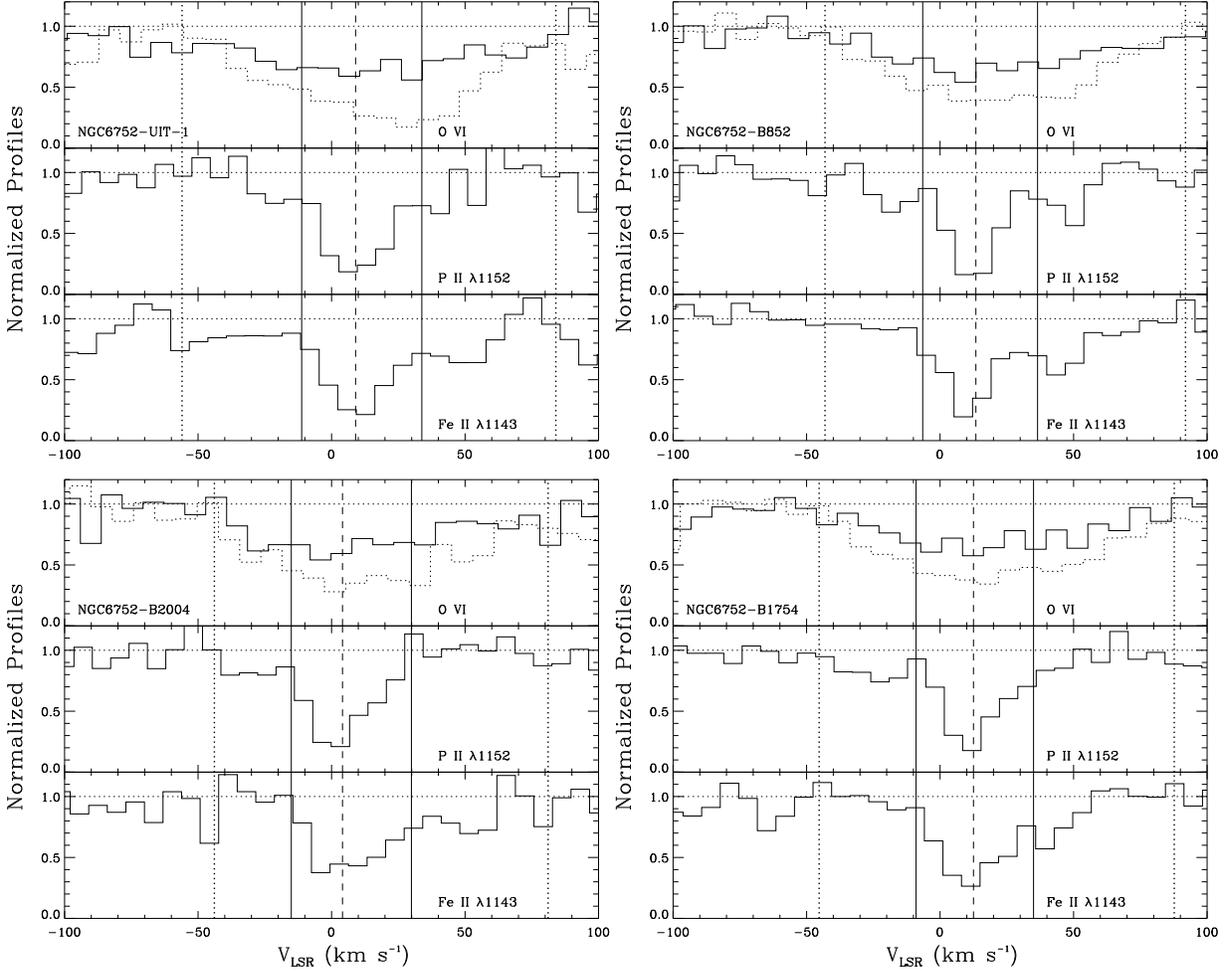}
\caption{Normalized profiles versus the velocity. For 
  O\,{\sc vi}, the lines at $\lambda$1032 (dotted line) and
  $\lambda$1038 (solid line) are shown. The vertical dotted lines
  indicate the extent of the O\,{\sc vi} absorption profile. The
  vertical solid lines indicate the low ions extent (anything outside
  this two lines are stellar lines). The dashed line represent the
  centroid of low ions; the typical uncertainty on this point is $\sim \pm
  8$ \km\ (including the statistic and systematic uncertainties).
\label{norm}}
\end{figure}

The reference frame for the {\fuse}\/ wavelength is heliocentric.
Toward this direction $v_{\rm helio}\simeq v_{\rm LSR}$, where $
v_{\rm LSR}$ is the local standard of rest (LSR) velocity. Since the
{\fuse}\/ absolute wavelength zero point is quite uncertain, we use
the known radial velocity of NGC\,6752 at $-24.4$ \km\ 
\citep{harris96}, and we assume that the 4 observed stars have this
radial velocity.  We use identified and uncontaminated stellar lines
(\ion{P}{4} $\lambda\lambda$1030.515, 1033.112) to correct the
velocity.  In Figure~\ref{norm}, we show the normalized spectra (see
\S~\ref{ana}) against the LSR velocity. There is a good agreement of
the cloud velocity toward the 4 stars. The mean centroid velocity for
\ion{P}{2} and \ion{Fe}{2} interstellar clouds is $ +9.8 \pm 4.2 \pm
5$ \km, where $\pm 4.2$ \km\ is the dispersion around the mean, and
$\pm 5$ \km\ the adopted relative velocity accuracy.  We note,
however, that the rotation curve of \citet{clemens85} for the
direction of NGC\,6752 predicts LSR radial velocities between $-45$ to
$0$ \km\ for clouds at distance $d \la 4$ kpc.

\section{Analysis}
\label{ana}

\subsection{Continuum Placement}

While the continuum placement can be very difficult to determine in
early-type stars and can remain uncertain especially near the
interstellar \ion{O}{6} absorption because of the strong \ion{O}{6}
winds in those stars \citep{lehner01,howk02b,zsargo03}, this is not a
concern for post-EHB stars.  The calibrated spectra of our four
targets in the region of the interstellar \ovi\ doublet are shown in
Figure \ref{spec}.  We identify the main absorption lines from atomic
and ionic species as well as molecular hydrogen above the top
spectrum.  All unmarked absorption lines are photospheric.
NGC\,6752-UIT-1 obviously shows the most metal lines in our sample,
and these will be a source of contamination for our \novi\ 
measurements.  The dotted lines show our adopted continua, modeled as
first order Legendre polynomials, are very good matches to the
line-free regions of these spectra.

\subsection{Column Density Measurements}

\begin{figure}[tbp]
\epsscale{0.8} 
\plotone{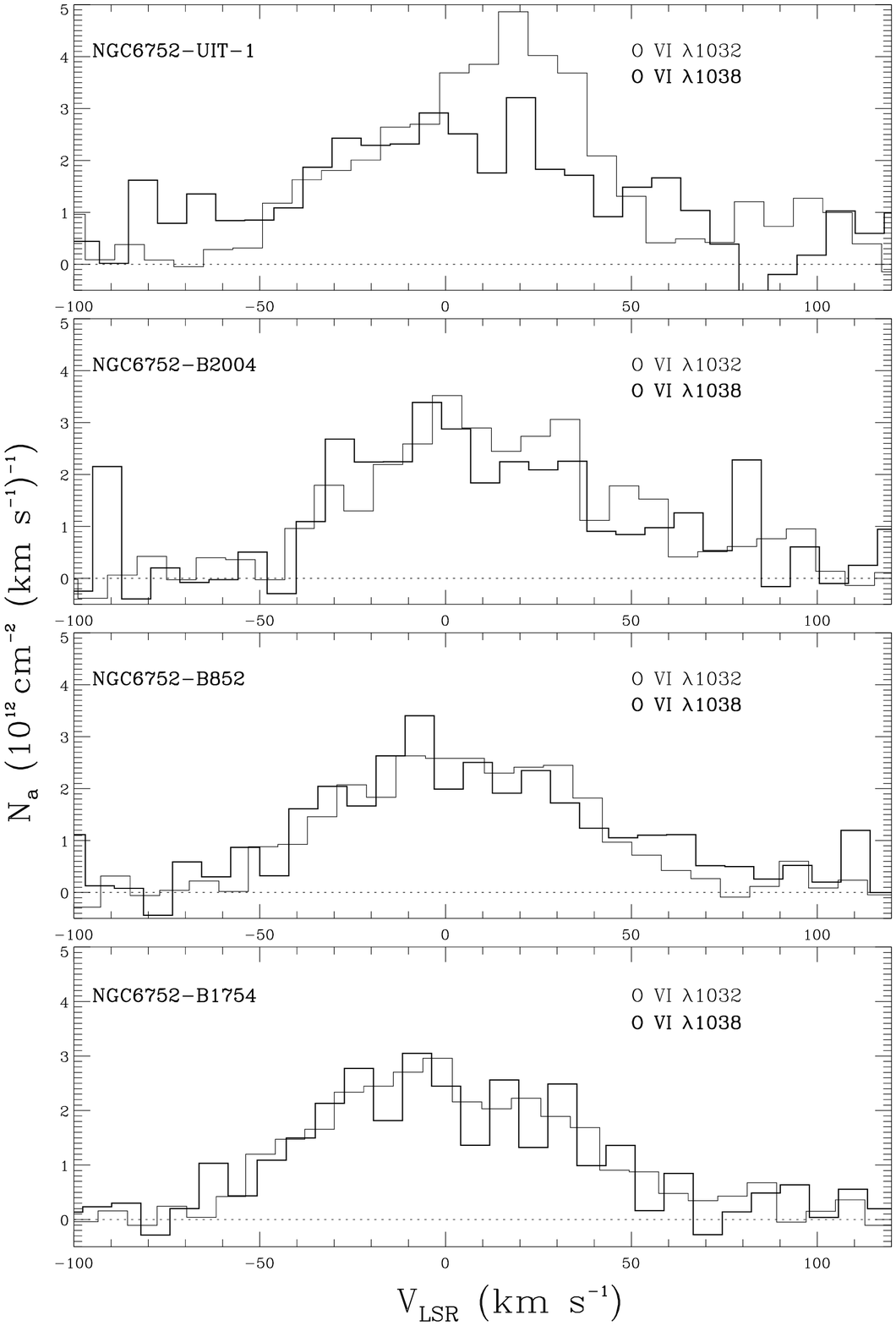}
\caption{Comparison of the apparent column density of O\,{\sc vi} 
  $\lambda$1032 and O\,{\sc vi} $\lambda$1038 for each sight lines.
\label{aod}}
\end{figure}

The \ion{O}{6} absorption line is broad enough to be fully resolved by
{\fuse}. We therefore use the apparent optical depth described by
\citet{savage91}.  In this method, the absorption profiles are
converted into apparent optical depth (AOD) per unit velocity,
$\tau_a(v) = \ln[I_{\rm c}/I_{\rm obs}(v)]$, where $I_{\rm obs}$,
$I_{\rm c}$ are the intensity with and without the absorption,
respectively.  The AOD, $\tau_a(v)$, is related to the apparent column
density per unit velocity, $N_a(v)$, through the relation
\begin{equation}
\label{col}
  N_a(v) = 3.768 \times 10^{14} 
     \frac{\tau_a(v)}{f \lambda(\mbox{\AA})} \,\, 
               {\rm cm}^{-2}\,({\rm km\,s^{-1}})^{-1}.
\end{equation}
The integrated apparent column density is equivalent to the true
integrated column density when the lines are fully resolved. In
Figure~\ref{aod}, we show the apparent column density comparison of
the \ion{O}{6} doublet for our four target stars. The $N_a(v)$
profiles for NGC\,6752-UIT-1 indicates some blending of \ion{O}{6}
$\lambda$1032, possibly with a stellar absorption line 
at $+ 15$ \km. Except for this
sight line, there is an excellent agreement between $N_a(v)$ for the
two \ion{O}{6} absorption lines, implying that the two lines contain
no unresolved saturated structure and that there is no contamination
from stellar lines or the HD 6--0 R(0) absorption line at 1031.912
\AA.  (We nevertheless checked that the HD line at 1066.271 \AA\ was
not present in our spectra to insure our \ovi\ profiles were not
contaminated by HD.)

The total column density is obtained by integrating the profile, $N =
\int N_a(v) dv $, between the velocity range $[-v,+v]$.  The results
are shown in Table~\ref{t2}, and again one can see the very good
agreement between the column density from \ion{O}{6} $\lambda$1032 and
\ion{O}{6} $\lambda$1038, except for \ion{O}{6} $\lambda$1032 toward
NGC\,6752-UIT-1, which is too large. In Table~\ref{t3}, the adopted
average column densities of the two \ion{O}{6} lines are indicated
(with the value for NGC\,6752-UIT-1 from \ion{O}{6} $\lambda$1038
only).

For comparison between the highly ionized gas and neutral and weakly
ionized gas, we also measure the column densities of \ion{P}{2} and
\ion{Fe}{2}. Many accessible lines exist for \ion{Fe}{2}, but several
are contaminated by stellar lines. We selected the weakest \ion{Fe}{2}
absorption lines that give consistent apparent column densities. Those
lines are \ion{Fe}{2} $\lambda$$\lambda$1055, 1112, and 1143. The
choice of these line spans different $f \lambda$, allowing us to check
that the lines are not saturated. Within $1\sigma$ error, the apparent
column densities of these lines are in agreement.  For \ion{P}{2}, we
have only one transition at 1152.818 \AA, but recent work by
\citet{lehner04} shows that for typically similar velocity
distribution and Galactic halo gas probed, this feature suffers little
or no saturation.

The quoted uncertainties are $1\sigma$.  They contain both statistical
(photon noise) and systematic (fixed-pattern noise, continuum
placement, velocity range over which the interstellar absorption lines
were integrated) contributions added in quadrature.

\subsection{Average Velocity and Velocity Dispersion}
The average line centroids and the velocity dispersions of \ion{O}{6}
are summarized in Table~\ref{t3}. They were estimated using both
\ion{O}{6} absorption lines and were computed with the following
expressions \citep{sembach92}:

\begin{equation}\label{vel}
\bar{v} = \frac{\int v N_a(v) dv}{\int N_a(v)dv} \,\, {\rm km\,s^{-1}}\,, 
\end{equation}

\begin{equation}\label{width}
b = \sqrt{\frac{2 \int (v -\bar{v})^2 N_a(v) dv}{\int N_a(v)dv}} \,\, {\rm km\,s^{-1}}.
\end{equation}

The quantity $b$ is related to the Doppler parameter used in the Voigt
profile fitting of interstellar lines. It contains all the
instrumental effects, thermal and turbulent broadening, and is
affected by the presence of multiple components.

The $1\sigma$ errors listed in Table~\ref{t3} include statistical
(photon counting) and systematic (continuum placement, velocity
cutoff) contributions.  We have restricted our measurements of the
velocity to the LiF\,1 channel to avoid spurious velocity shifts that
may exist between LiF\,1 and LiF\,2.  The second error of $\pm 5$ \km\ 
for $\bar{v}$ is the systematic uncertainty from the absolute
wavelength scale (see \S~\ref{obs}).

The average line centroid of \ion{O}{6}, $+16.0 \pm 5.3 \pm 5$ \km, is
$\sim +6$ \km\ higher than the average velocity of \ion{P}{2} and
\ion{Fe}{2}, but consistent within $1\sigma$ errors (see
Figure~\ref{spec}).


\section{Discussion: \ovi\ Small-Scale Structure}
\label{disc}

\subsection{Small-Scale Structure Toward NGC 6752}

\begin{figure}[tbp]
\epsscale{1} 
\plotone{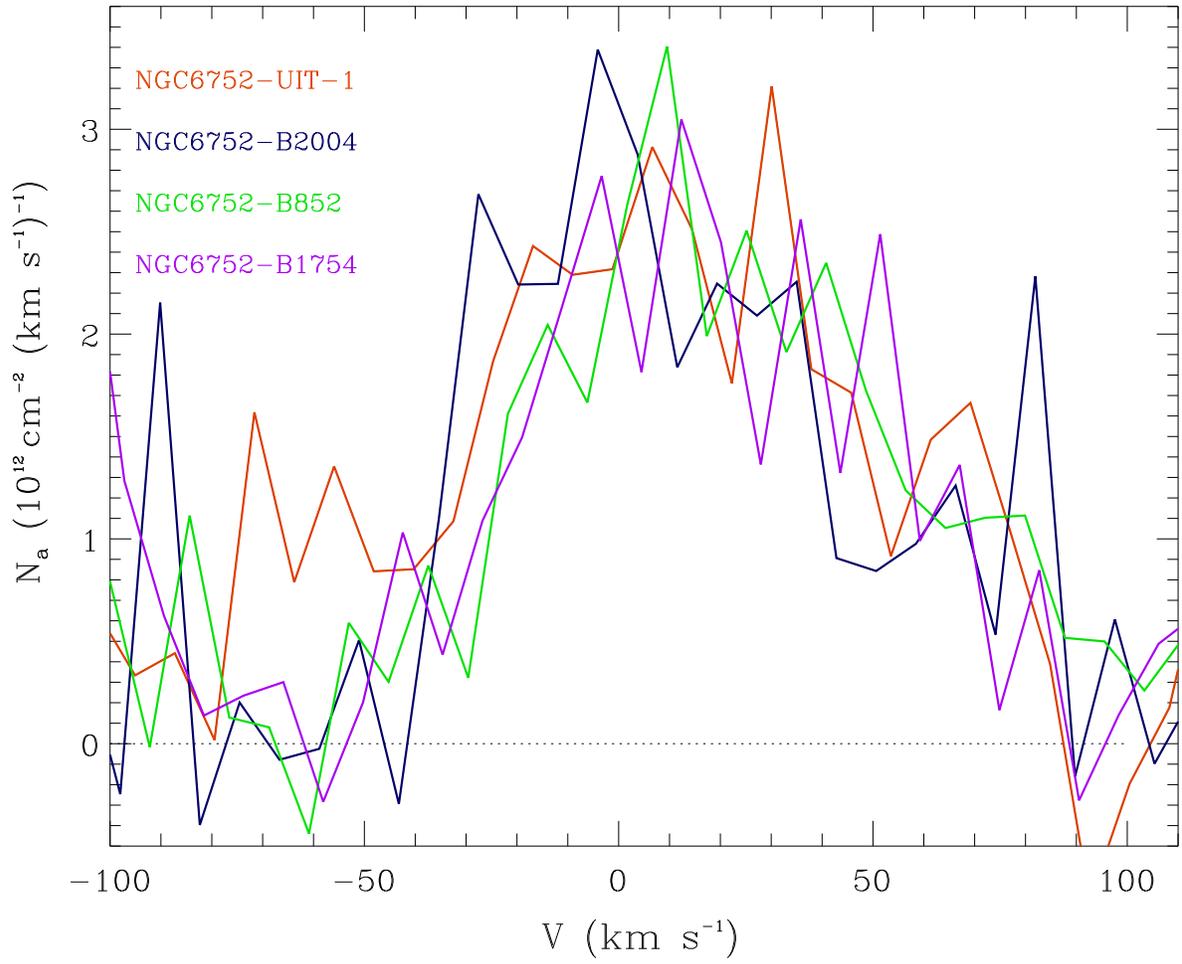}
\caption{Comparison of the apparent column density of O\,{\sc vi} 
  $\lambda$1038 toward each sight lines.
\label{aod1}}
\end{figure}

We have presented measurements of \ovi\ column densities toward four
stars in the globular cluster NGC\,6752.  There is good agreement of
the \ion{O}{6} column densities, the average line centroids, and the
velocity dispersions of \ion{O}{6} along these four sight lines. 
This is further shown in Figure~\ref{aod1} where  the apparent column density 
of \ion{O}{6} $\lambda$1038 toward each sight lines is
compared.   The
sight lines in our study are separated by angular distances of
$2\arcmin \la \Delta \theta \la 9 \arcmin$, corresponding to spatial
separations of $\sim2$ pc to $\sim9$ pc at the distance of the
globular cluster.  The dispersion about the mean column density in
this direction is only $\pm0.03$ dex, or $1.4\times10^{13}$ \column.
We can limit column density variations between these sight lines to
$\Delta \N{\ovi} \la 8\times10^{13}$ \column\ ($2\sigma$), or $\Delta
\N{\ovi} \la 4\times10^{13}$ \column\ ($2\sigma$) if we exclude
NGC\,6752-UIT-1 (in case weak stellar lines contaminate the \ovi\ 
1037.617 \AA\ transition).  

Our data also show the kinematics of the absorption along each of
these sight lines to be indistinguishable.  In short, there is no
evidence that we are probing any differences in the properties of the
\ovi -bearing gas as we compare these four sight lines.  Neither the
column densities nor the kinematics of \ovi\ toward NGC\,6752 are
peculiar with respect to other sight lines at similar galactic
latitudes \citep[see][]{savage03}.

The low ions (\ion{P}{2} and \ion{Fe}{2}) also show no evidence for
column density variations among the four sight lines studied in this
work.  Concerns about unresolved saturation in these profiles makes
the limits on relative variations in these species larger (i.e., less
constraining).  The \ion{P}{2} and \ion{Fe}{2} are significantly
narrower than the \ovi\ profiles along these sight lines.  This is
likely in part due to the different temperatures of gas probed by
these species.  However, it should be noted that the profiles of the
strongest low-ion absorption lines in the \fuse\ data, such as the
strongly-saturated \ion{C}{2} 1036.337 \AA\ transition, have the same
total breadth (defined as the approximate velocities at which the
absorption profiles return to the continuum) as \ovi.  This could
imply that the \ovi\ absorption is arising in interfaces between
hotter and cooler gas, with the total amount of \ovi\ as a function of
velocity indicating the relative number of interfaces in each velocity
range.

Absorption from the low ions is clearly (and expectedly) tracing
different material along these sight lines, given the very different
kinematic properties compared to the \ovi\ absorption. The \ion{O}{6}
profiles are clearly more extended than the profiles of the low ions
(see Figure~\ref{norm}), implying that \ion{O}{6} absorption contains
more and/or broader components than the low ions.

While we see no evidence for variations in \ovi\ column densities over
the $\la9\arcmin$ angular scales probed by the current observations,
there is evidence for variations over larger scales in this region of
the sky. \citet{savage03} report on \fuse\ observations of the AGN ESO
141-G55, some $1\fdg8$ from the center of NGC 6752.  They derive an
\ion{O}{6} column density of $\log \novi = 14.50\pm 0.02 \pm 0.02$
(random and systematic uncertainties, respectively) toward
ESO\,141-G55, +0.16 dex larger than toward NGC\,6752.  This could be a
signature of significant \ovi\ at distances greater than that of NGC
6752; indeed, if the scale-height of \ovi\ is $\sim2.3$ kpc (Savage et
al. 2003), then this cluster may only reside behind $\sim50$\% of the
total \ovi\ column density in this direction.

\subsection{Comparison with the LMC and SMC Lines of Sight}

\citet{howk02a} presented observations of Galactic \ovi\ absorption
toward 23 stars in the Magellanic Clouds drawn from the sight lines
studied by \citet{howk02b} (for the LMC) and \citet{hoopes02} (for the
SMC).  The angular separations of the observed stars range from
$\sim3\farcm0$ to $\sim5\fdg0$.  Most of the sight lines studied by
Howk et al. in the LMC were separated by much larger angular scales
than those toward NGC 6752, with a minimum angular separation of
$\approx30\arcmin$.  The angular separations of the sample of stars
studied in the SMC is closer to those in the current sample.  The
smallest separations studied toward the SMC were $3\farcm0$,
$9\farcm1$, $10\farcm5$.  Thus, the smallest separations were on
scales similar to those studied in this work.  Assuming the bulk of
the \ovi\ in these directions resides within two vertical scale
heights, i.e., $z_{O\, VI} \la 2 h_z ~ 5.0$ kpc (Savage et al. 2003),
the corresponding spatial scales probed by these observations are $\ll
80$--800 pc toward the LMC and $\ll 6$--400 pc \citep{howk02a}.

\ovi\ column density variations were seen over effectively all scales
probed.  Typical variations were in the range $\sim25$\% to
$\sim50$\%, although there were pairings of sight lines with much
smaller variations (consistent with no change) and larger variations
(up to a factor of 4 variation between sight lines separated by
$1^\circ$ to $2^\circ$).  \citet{howk02a} discuss the angular
correlation function of the \ovi\ measurements, finding no preferred
angular scale for the \ovi\ variations.  \citet{savage03} showed that
the variations seen toward the LMC and SMC extend to larger angular
scales.  On smaller angular scales, \citet{danforth02} have reported a
lack of column density variations greater than $\sim12$\% toward
several stars in the SMC cluster NGC\,346 with an average separation
of $\la19\arcsec$.  Thus, the SMC observations suggest the \ovi\ is
smooth on scales less than an arcminute and patchy on slightly larger
scales.

The lack of \ovi\ column density variations in the current study on
scales as small as those identified toward the SMC, may be due to the
different sight lines these two samples probe through the Galactic
ISM.  However, there are two concerns that should be mentioned
regarding the Magellanic Cloud measurements that are not existing
toward NGC\,6752, and both of which are probably more extreme for the
SMC sight lines.  First, the stellar continua in the region of
interstellar \ovi\ are quite complex for the early-type stars used to
probe the Magellanic Clouds.  \citet{howk02b} and \citet{hoopes02}
discuss in detail the difficulties of accurately determining a
continuum in the face of these complexities.  The variations claimed
over angular scale of $3\farcm0$ by \citet{howk02a} are potentially
susceptible to systematic uncertainties associated with the placement
of the stellar continuum.  We believe the SMC measurements would be
more likely to suffer from this source of systematic error given the
stellar winds that dominate the shape of the stellar continuum in this
wavelength region are significantly better developed and well behaved
in the LMC.  Indeed, among the closest pairing of stars in the SMC,
the direction toward Av\,83 has a much smaller \ovi\ column density
(13.93 dex) than toward two nearby directions (Av\,75 and Av\,95, with
$\log \novi = 14.15$ and 14.14, respectively), but the stellar
continua of these stars are among the least certain in the sample of
\citet{hoopes02}.  The continuum for Av\,83 is especially difficult.
If the latter star is removed from the sample, the smallest scale in
\citet{howk02a} over which \ovi\ variations are seen would be $ \sim
10\arcmin$.  Reexamining the data for Av\,83, we favor this scenario.
The problems associated with stellar continuum placement uncertainties
are not applicable to the NGC\,6752 sight lines given the nature of
the observed stars (see Figure~\ref{spec}).

Another possible source of systematic error for Milky Way measurements
in the Magellanic Cloud directions is the presence of material flowing
out of these galaxies with velocities that overlap the expected Milky
Way velocities or intermediate-velocity gas not well separated from
the Galactic halo gas \citep{hoopes01,lehner01a,danforth03}.  For the
SMC the outflow velocities required are $\sim150$ \kms; for the LMC,
the outflow velocities required are $\ga 200$ \kms.  Thus, it seems
more likely that this effect would contaminate directions toward the
SMC.  \citet{hoopes01} have reported on the detection of the receding
side of a supernova remnant in \ovi\ toward the SMC star HD 5980.  The
analysis of \citet{danforth03} suggested that the approaching side of
the shell may overlap the low-velocity Milky Way \ovi\ toward stars
projected onto the remnant.  Their measurements showed the Milky Way
\ovi\ to be stronger in directions toward the remnant than those
several arcminutes outside the remnant.  Unfortunately this result
could imply either that arcminute variations in the Galactic halo
\ovi\ are seen in this direction, or that the approaching side of the
remnant is contaminating the Milky Way gas. 

It is tempting to repeat the analysis of the \ovi\ angular correlation
function discussed by Howk et al., including the current measurements.
However, the likelihood that the sight line to NGC 6752 only probes a
fraction of the total \ovi\ integrated through the Galaxy, while the
Howk et al. data extend through the entire halo, makes this seem like
a somewhat dubious endeavor.


\section{Concluding Remarks}
\label{sum}

We have presented observations of Galactic halo \ovi\ absorption
toward 4 post-EHB stars in NGC\,6752 situated at a distance of 3.9 kpc
and separated by only $2\farcm2 - 8\farcm9$ allowing to study the hot
interstellar gas on small spatial scales of $\la 2.5-10.1$ pc. The
continuum of these stars is well behaved and we were able to derive
accurate \ovi\ measurements.

A comparison of the present results with the results of
\citet{howk02a} toward the LMC and SMC and the results of
\citet{danforth02} toward the SMC cluster NGC\,346 shows that there is
a higher frequency of \novi\ variations on degree-scale separation
greater than $\Delta \theta \ga 30\arcmin $.  On degree-scale
separation smaller than $\Delta \theta \la 12\arcmin $ corresponding
to spatial-scale separation $\la 10$ pc, \ion{O}{6} is more smoothly
distributed and does not have the complexity of the larger-scale
distribution.

\citet{howk02a} and \citet{savage03} discussed the existence of large
column density variations over various angular scale favored models in
which the \ovi\--bearing medium is composed of small complex
cloud-like or sheet-like distributions of material (such envisioned in
various interface models), rather than models in which \ovi\ is
distributed in a diffuse, smooth medium.  If \ovi\ is in small complex
clouds, these clouds cannot generally be smaller than 10 pc in the
direction of NGC 6752 unless we are seeing these structures in a
face-on orientation.

\acknowledgments

This program was made possible by support from NASA under award
NAG5-12345.  We thank W. Landsman for making available the UIT data
used in making Figure 1.  We thank our collaborators W. Landsman and
S. Moehler for their input in planning these observations.

\newpage

\begin{deluxetable}{lcccccccl}
\tablecolumns{9}
\tablewidth{0pc} 
\tabletypesize{\footnotesize}
\tablecaption{Observation Summary \label{t1}} 
\tablehead{\colhead{Object}    & \colhead{R.A.}    &   \colhead{Dec.}&\colhead{$l$} &\colhead{$b$}&\colhead{$t_{\rm exp}$} &   \colhead{S/N}&   \colhead{ID}&   \colhead{Observation}     \\
\colhead{}&\colhead{(h\ m\ s)}&\colhead{($\circ$\  '\  ")}&\colhead{($\circ$)}&\colhead{($\circ$)}&   \colhead{(ks)}&   \colhead{}&   \colhead{}&   \colhead{Date}}
\startdata
NGC\,6752-UIT-1         & 19 10 54.50      &    $-$59 59 46.40  & 336.48  &$-$25.64 & 30.3& 11  &  C0760101 & Apr. 10 2002   \\
NGC\,6752-B2004         & 19 11 04.78      &    $-$59 57 59.19  & 336.52  &$-$25.65 & 38.1& 10  &  C0760201 & Jun. 04 2002   \\
NGC\,6752-B852          & 19 11 27.89      &    $-$60 00 39.10  & 336.48  &$-$25.71 & 34.9& 12  &  C0760301 & Apr. 14 2002 \\
NGC\,6752-B1754         & 19 11 08.87      &    $-$59 52 20.70  & 336.62  &$-$25.64 & 36.7& 14  &  C0760401 & Apr. 15 2002  \\
\enddata
\tablecomments{Typical signal-to-noise (S/N) may fluctuate slightly
  depending on which wavelength range is studied.  The S/N is per
  $\sim$27 m\AA\ binned pixel.}
\end{deluxetable}

\begin{deluxetable}{lcccc}
\tablecolumns{5}
\tablewidth{0pc} 
\tabletypesize{\footnotesize}
\tablecaption{Colum Densities \label{t2}} 
\tablehead{\colhead{Ions}    & \colhead{NGC\,6752-UIT-1}    &   \colhead{NGC\,6752-B2004}&\colhead{NGC\,6752-B852} &\colhead{NGC\,6752-B1754} \\
           \colhead{}    & \colhead{$\log N$ $[-v,+v]$}    &   \colhead{$\log N$ $[-v,+v]$}&\colhead{$\log N$ $[-v,+v]$} & \colhead{$\log N$ $[-v,+v]$} \\
           \colhead{}    & \colhead{(cm$^{-2}$, \km)}    &   \colhead{(cm$^{-2}$, \km)}&\colhead{(cm$^{-2}$, \km)} & \colhead{(cm$^{-2}$, \km)}}
\startdata
O\,{\sc vi}  $\lambda$1032 & $14.47 \pm 0.04$ $[-56, +84]$  & $14.36 \pm 0.05$ $[-44, +68]$  & $14.33 \pm 0.04$  $[-43, +92]$ & $14.34 \pm 0.02$  $[-45, +88]$  \\
O\,{\sc vi}  $\lambda$1038 & $14.39 \pm 0.06$ $[-56, +84]$  & $14.34 \pm 0.06$ $[-44, +68]$  & $14.35 \pm 0.04$  $[-43, +92]$ & $14.33 \pm 0.03$  $[-45, +88]$  \\
P\,{\sc ii}  $\lambda$1152 & $13.75 \pm 0.06$ $[-11, +34]$  & $13.66 \pm 0.10$ $[-15, +30]$  & $13.65 \pm 0.04$  $[-6, +36]$ & $13.66 \pm 0.04$  $[-9, +35]$  \\
Fe\,{\sc ii}               & $14.80 \pm 0.07$ $[-11, +34]$  & $14.82 \pm 0.10$ $[-15, +30]$  & $14.80 \pm 0.07$  $[-6, +36]$ & $14.80 \pm 0.02$  $[-9, +35]$  
\enddata
\tablecomments{The Fe\,{\sc ii} measurements result from 3 absorption lines at 1055 \AA, 1112 \AA, and 1143 \AA. 
Toward NGC\,6752-UIT-1, O\,{\sc vi}  $\lambda$1032 is contaminated by a stellar line, and another stellar
line possibly contaminates also P\,{\sc ii}  $\lambda$1152. 
}
\end{deluxetable}

\begin{deluxetable}{lcccc}
\tablecolumns{5}
\tablewidth{0pc} 
\tabletypesize{\footnotesize}
\tablecaption{Adopted Column Densities, Line Widths, and Velocities of O\,{\sc vi} Absorption \label{t3}} 
\tablehead{\colhead{Target} & \colhead{$\log N$} & \colhead{$\log (N \sin \mid\! b\mid\!)$}     & \colhead{$b$} &\colhead{$\bar{v}_{\rm LSR}$}  \\
           \colhead{}    & \colhead{(dex)}  & \colhead{(dex)}   &   \colhead{(\km)}&\colhead{(\km)} }
\startdata
NGC\,6752-UIT-1 & $14.39 \pm 0.06$ & $14.03 \pm 0.06$ & $42.4 \pm 4.5$ & $+11.8 \pm 5.8 \pm 5.0 $ \\     
NGC\,6752-B2004 & $14.35 \pm 0.04$ & $13.99 \pm 0.04$ & $42.4 \pm 2.2$ & $+11.3\pm 3.0 \pm 5.0 $ \\      
NGC\,6752-B852  & $14.34 \pm 0.03$ & $13.98 \pm 0.03$ & $43.3 \pm 1.6$ & $+22.0 \pm 2.7 \pm 5.0 $ \\     
NGC\,6752-B1754 & $14.34 \pm 0.02$ & $13.98 \pm 0.02$ & $42.5 \pm 1.1$ & $+18.9 \pm 2.0 \pm 5.0 $ 
\enddata
\end{deluxetable}

\clearpage

\end{document}